\documentclass[apj,twocolumn]{emulateapj}

\slugcomment{Submitted to ApJ}

\shortauthors{Parmentier}
\shorttitle{Star Formation Relations}

\newcommand\eff{\epsilon_{\rm ff}}
\newcommand\tff{\tau_{\rm ff}}

\newcommand\Ls{L_{\odot}}
\newcommand\Ms{M_{\odot}}

\newcommand\Mspp{M_{\odot} \cdot pc^{-2}}
\newcommand\Msppp{M_{\odot} \cdot pc^{-3}}
\newcommand\uHCN{\rm K \cdot km \cdot s^{-1} \cdot pc^2}

\newcommand\lhcn{L_{\rm HCN}}
\newcommand\lir{L_{\rm IR}}
\newcommand\nhh{n_{\rm H2}}

\newcommand\fft{free-fall time }
\newcommand\sfe{star formation efficiency }
\newcommand\sfr{star formation rate }
\newcommand\stf{star formation }
\newcommand\sfing {star-forming }

\newcommand\Son{Solar neighborhood }

\begin{document}



\title{The dense-gas mass versus star formation rate relation: \\ a misleading linearity?}


\author{G.~Parmentier\altaffilmark{1}}


\altaffiltext{1}{Astronomisches Rechen-Institut, Zentrum f\"ur Astronomie der Universit\"at Heidelberg, M\"onchhofstr. 12-14, D-69120 Heidelberg, Germany}


\begin{abstract}

We model the star formation relation of molecular clumps in dependence of their dense-gas mass when their volume density profile is that of an isothermal sphere, i.e. $\rho_{clump}(r) \propto r^{-2}$.  Dense gas is defined as gas whose volume density is higher than a threshold $\rho_{th}=700\,\Msppp$, i.e. HCN(1-0)-mapped gas.  We divide the clump into two regions: a dense inner region (where $\rho_{clump}(r) \geq \rho_{th}$), and low-density outskirts (where $\rho_{clump}(r) < \rho_{th}$).  We find that the total star formation rate of clumps scales linearly with the mass of their dense inner region, even when more than half of the clump \stf activity takes place in the low-density outskirts.  We therefore emphasize that a linear star formation relation does not necessarily imply that \stf takes place exclusively in the gas whose mass is given by the \stf relation.  The linearity of the \stf relation is strengthened when we account for the mass of dense fragments (e.g. cores, fibers) seeding \stf in the low-density outskirts, and which our adopted clump density profile $\rho_{clump}(r)$ does not resolve.  We also find that the star formation relation is significantly tighter when considering the dense gas than when considering all the clump gas, as observed for molecular clouds of the Galactic plane.  When the clumps have no low-density outskirts (i.e. they consist of dense gas only), the star formation relation becomes superlinear and progressively wider.
  
\end{abstract}



\section{Introduction}
\label{sec:intro}
Star-forming molecular clumps of the Galactic disk show a linear correlation between their HCN(1-0) emission, $\lhcn$, and their total infrared luminosity, $\lir$, when $\lir > 10^{4.5}\Ls $ \citep{wu05}. 
The $\lhcn - \lir$ correlation for molecular clumps extends that observed by \citet{gs04} for galaxies, from spirals to luminous and ultraluminous infrared galaxies (hereafter LIRGS and ULIRGS, respectively).  That is, over eight orders of magnitude in luminosity, from molecular clumps up to ULIRGS, the infrared and HCN luminosities appear linearly correlated.  

The HCN line emission traces dense molecular gas, that is, gas with a mean number density $\nhh \simeq 3 \cdot 10^4 cm^{-3}$.  The total infrared luminosity provides a tracer of the \stf rate when \stf takes place in optically-thick gas (i.e. gas optically thick to stellar UV photons, where the stellar optical light is absorbed and re-radiated in the infrared).  The observed $\lhcn - \lir$ correlation thus constitutes the imprint of a linear relation between the dense gas mass and the \sfr of Galactic clumps and galaxies.  Additionally, the correlation found for galaxies by \citet{gs04} may also extend to the dense-gas content and \stf activity of molecular clouds of the Solar neighbourhood \citep{lad12}.  A caveat here is that \citet{gs04} and \citet{lad12} do not define the dense gas in the same way.  \citet{lad12} identify the dense gas of nearby clouds as gas with an extinction $A_K \geq 0.8\,$mag, equivalent to a mass surface density $\Sigma_{gas} \geq 160\,\Mspp$.  HCN surveys of nearby molecular clouds do not exist yet, and it is unclear how well the dense gas masses derived from such surveys would agree with those based on an extinction criterion as done by \citet{lad12}.  

Adding to the results of \citet{wu05} and \citet{lad12}, \citet{vut16} find that, when summing up the \sfr and the dense gas mass of Galactic molecular clouds, the ratio of both quantities is equal to that obtained for ULIRGs within the error bars.  Their result stands for both nearby clouds, where the dense gas is defined as gas with $A_{\rm V} \geq 8$\,mag \citep{eva14}, and for more distant molecular clouds (heliocentric distance ranging from $\simeq 2$ to $13$\,kpc) whose dense gas is mapped using 1.1mm dust emission from the Bolocam Galactic Plane Survey \citep{agu11, gin13}.   

All these results have led to the proposal that HCN molecular clumps are the basic units of star formation in galaxies
\citep{wu05} or, more generally, that the bulk of star formation takes place in molecular gas whose surface or volume density is higher than some threshold value \citep{wu10,lad10,hei10}.  This in turn has led to applications in the fields of star cluster systems \citep{par11a} and massive-star formation \citep{par11b}.
It is interesting to note, however,  that \citet{lad12} and \citet{vut16} interpret the dense-gas star-formation-rate correlation as the \sfr being {\it controlled}, or {\it predicted}, by the dense-gas mass of the star-forming region.  As we shall see in this contribution, one has indeed to distinguish between {\it (i)} using the dense-gas mass as a predictor of the \stf rate, and {\it (ii)} arguing that only cluster-forming regions with a mean number density higher than a few $10^4\,{\rm cm}^{-3}$ give rise to star formation.  

On the scale of galaxies, the $\lhcn-\lir$ linear correlation of \citet{gs04} constitutes most of the observational support for star-formation models implementing a volume density threshold for star formation.  Yet, \citet{kru07b} show that a linear $\lhcn - \lir$ correlation may also stem from the median density of the observed molecular clouds being smaller than the critical density of the molecular line, here HCN.  Once the cloud median density becomes higher than the line critical density, the linear correlation breaks down and the correlation becomes superlinear.  \citet{gao07} indeed observe an upturn in the $\lhcn-\lir$ correlation at high infrared luminosity.  In addition, \citet{gb12} find a superlinear relation between the HCN and infrared luminosities of star-forming regions (size 1.7-3.6\,kpc) in a sample of normal galaxies, LIRGs and ULIRGs \citep[see also][]{gc08}.  
  
Steeper-than-linear \stf relations provide hints for a more complex \stf picture than a mere molecular-gas volume-density threshold. 

Observations in the \Son and the Large Magellanic Cloud actually show that \stf also takes place in low-density environments, that is, regions where the number density averaged on a pc-scale is lower than $10^4\,cm^{-3}$.  For instance, LH95 is a star-forming region of the Large Magellanic Cloud which currently forms three subclusters, each with a mean stellar density $\gtrsim 1\,\Msppp$ on spatial scales of a few pc \citep[][his table 2]{dar09}.  Even assuming that the \sfe of these subclusters is as low as 0.1, the mean volume density of the gas that has given rise to them is still about 2 orders of magnitude lower than that of HCN clumps.  

Molecular clouds of the \Son are exquisite targets to understand how star and gas properties relate to each other, given that their Young Stellar Objects (hereafter YSOs) can be counted \citep[e.g.][]{eva09, gut11}.  Their observation by {\it Spitzer} and {\it Herschel} has revealed a power-law relation between the surface densities of gas and YSOs, when the densities are defined locally, i.e. at the location of each YSO \citep{gut11}, or in gas-surface-density contours \citep{hei10,lad12,lom13,eva14}.  Its slope varies from one study to another, but is markedly superlinear and close to 2.  \citet{par13} show that a quadratic relation between the local surface densities of gas and YSOs is recovered if \stf takes place in centrally-concentrated spherical clumps with a constant \sfe per free-fall time, $\eff$.  Their model does not hinge on a gas volume-density threshold: star formation does take place in low-density clumps, or in the low-density outskirts of clumps, albeit at a slow rate due to the long free-fall time there.  Here we caution that the gas volume densities considered in \citet{par13}, $\rho_{gas}$, are averaged over spatial scales larger than the size of the dense fragments/cores in which individual stars form.  That is, their clump density profiles do not resolve the density peaks associated to such fragments/cores.        

Figures~1 and 7 of \citet{par13} present the volume density profiles and the cumulated masses, respectively, of gas and stars in a spherical clump with an initial gas mass of $10^4\,\Ms$ and a radius of 6\,pc (i.e. a clump mean density of $12\,\Msppp$ as can be found in cluster-forming regions of nearby molecular clouds).  
Inspecting those figures reveals an intriguing fact: only 1/3 of the total stellar mass forms in gas whose initial density is higher than $\nhh \simeq 10^4\,cm^{-3}$ (or $\rho_{gas} > 700\,\Msppp$).  Hence, even in models in which star formation is more efficient
in the dense inner region of a molecular clump, the bulk of \stf may still take place {\it outside} that region.
This raises one question: in such models, are the dense-gas mass and \sfr still linearly correlated, as is observed by \citet{wu05}?

The key message of this contribution will be the following: {\it (i)} if molecular clumps are the basic units of (clustered) star formation, {\it (ii)} if their initial radial density profile is that of an isothermal sphere, and {\it (iii)} if we divide each clump into an inner and an outer regions based on some volume density threshold, then the {\it total} \sfr of clumps correlates linearly with the mass of their dense inner region even if substantial \stf also takes place in the outer one.  

The outline of the paper is the following.  Section~\ref{sec:sfrmth} presents analytical insights.  In Section \ref{sec:sim}, simulations are carried out to map the stellar mass fraction formed in the dense inner region as a function of clump mass and radius.  We also discuss how the dense-gas mass, clump mass and \sfr relate to each other.  Conclusions are presented in Section \ref{sec:conclu}.

Unless otherwise quoted, the term `density' refers to the volume density.

\section{Analytical Insights}
\label{sec:sfrmth}

We assume that \stf is characterized by a constant \sfe per free-fall time, $\eff$ \citep{kru05}.  The \sfr of a gas reservoir of mass $m_{gas}$ is then given by:
\begin{equation}
SFR_{gas} \simeq \eff \frac{m_{gas}}{\tau_{ff,gas}}\,,
\label{eq:sfr_gas}
\end{equation}
with $\tau_{ff,gas}$ the free-fall time at the gas mean density:

\begin{equation}
\tau_{ff, gas} = \sqrt{ \frac{3\pi}{32G\rho_{gas}} }
\label{eq:tff_gas}
\end{equation}
($G$ is the gravitational constant).

\subsection{Star formation in dense gas implies a linear $L_{HCN}$-$L_{IR}$ correlation}
\label{ssec:dg}

A linear correlation between the mass of dense HCN-mapped gas and the \sfr of molecular clouds is expected if star formation is limited to their dense gas.  The \sfr of a dense clump whose \fft and mass are $\tau_{ff,dg}$ and $m_{dg}$, respectively, obeys:
\begin{equation}
SFR_{dg} = \zeta \eff \frac{m_{dg}}{\tau_{ff,dg}}\;.
\label{eq:sfr_dg}
\end{equation}
$\zeta$ is a factor accounting for the centrally-concentrated nature of the clump, which yields a global \sfe higher than $\eff$ after one free-fall time.  Its value is $\zeta \simeq 1.6$ \citep{tan06,par14a}.  

Given the limited range in volume densities of HCN-mapped gas, the free-fall time $\tau_{ff,dg}$ of HCN-clumps is about constant, which yields a linear scaling between their mass $m_{dg}$ and their \sfr $SFR_{dg}$.  

Using standard conversion factors to derive the clump infrared luminosity \citep{ken98b}:
\begin{equation}
SFR [\Ms \cdot Myr^{-1}] = 2 \cdot 10^{-4} L_{IR}^{\rm SFR} [\Ls]\,,
\label{eq:ken98}
\end{equation}
and HCN luminosity \citep{gs04}:
\begin{equation}
m_{dg} [\Ms] = 10 L'_{HCN} [\uHCN]\,,
\label{eq:dghcn}
\end{equation}
one infers the corresponding $L_{IR} - L'_{HCN}$ relation:
\begin{equation}
L_{IR}^{\rm SFR} [\Ls] = 5 \cdot 10^4 \frac{\zeta \eff}{\tau_{ff,dg}} L'_{HCN} \, [\uHCN]\,,
\label{eq:lirlhcn}
\end{equation}
with $\tau_{ff,dg}$ expressed in units of Myrs.
\citep[For a discussion of the conversion factors used in Eqs \ref{eq:ken98} and \ref{eq:dghcn}, see e.g. ][]{gb12}.

The \sfe per \fft can be estimated by comparing Eq.~\ref{eq:lirlhcn} with the result of \citet{wu05}, i.e. $log_{10}(\lir) = log_{10}(\lhcn) + 2.8$.  Prior to doing so, we multiply the infrared luminosities of \citet{wu05} by a factor of 2 given that they are underestimated due to beam-size limitations \citep{vut13}.  The median number density of the HCN clumps of \citet{wu05} with $L_{IR} > 10^{4.5}\,\Ls$ is $n_{\rm H2} \simeq 10^4\,cm^{-3}$ \citep[see top-right panel in fig.~35 of][]{wu10}, equivalent to a mass density $\langle\rho_{dg}\rangle \simeq 700\,\Msppp$ and a \fft $\tau_{ff,dg} \simeq 0.3$\,Myr.  
For Eq.~\ref{eq:lirlhcn} to match the updated relation of \citet{wu05}, one has to adopt a \sfe per \fft of 0.005.  

The actual value of $\eff$ is likely higher than $\eff = 0.005$, however.  This is because Eq.~\ref{eq:ken98} has been devised for galaxies, that is, \stf time-scales of 10-100\,Myr and a fully sampled stellar IMF, two conditions not necessarily met in individual molecular clumps.  For a given infrared luminosity, Eq.~\ref{eq:ken98} underestimates the true \sfr of molecular clumps and clouds by a factor of 3 to 10 \citep[see e.g.][]{kru07, hei10, urb10, lad12}.  In what follows we assume an underestimating factor of 5 and adopt a \sfe per \fft $\eff = 0.025$.

If the bulk of \stf takes place in dense molecular gas, an $L_{IR}$-$L_{HCN}$ correlation (Eq.~\ref{eq:lirlhcn}) is observed.  But is the converse always true?  That is, does $L_{IR} \propto L_{HCN}$ necessarily imply that cluster-forming regions have a mean density of order a few $10^4\,{\rm cm}^{-3}$?

\subsection{A $L_{HCN}$-$L_{IR}$ linear correlation does not necessarily imply cluster formation in dense gas only}
\label{ssec:no}

Let us consider a spherical clump of gas with a mass $m_{clump}$ and a radius $r_{clump}$, without making any assumption about its mean volume density.  Its \sfr is: 
\begin{equation}
SFR_{clump} = \zeta \eff \frac{m_{clump}}{\tff}\,,
\label{eq:sfrbasic}
\end{equation}
with $\tff$ the free-fall time at the clump mean volume density, i.e.:
\begin{equation}
\tff = \sqrt{ \frac{ 3 \pi }{ 32 G \rho_{clump} } } = \sqrt{ \frac{ \pi^2 \, r_{clump}^3}{ 8 G \, m_{clump} } }\;.
\end{equation}
Equation~\ref{eq:sfrbasic} can therefore be rewritten as:
\begin{equation}
SFR_{clump} = \zeta \eff \frac{\sqrt{8G}}{\pi} \left( \frac{m_{clump}}{r_{clump}} \right)^{3/2} \;.
\label{eq:sfrcl}
\end{equation}
 
\begin{figure}
\begin{center}
\epsscale{1.1}  
\plotone{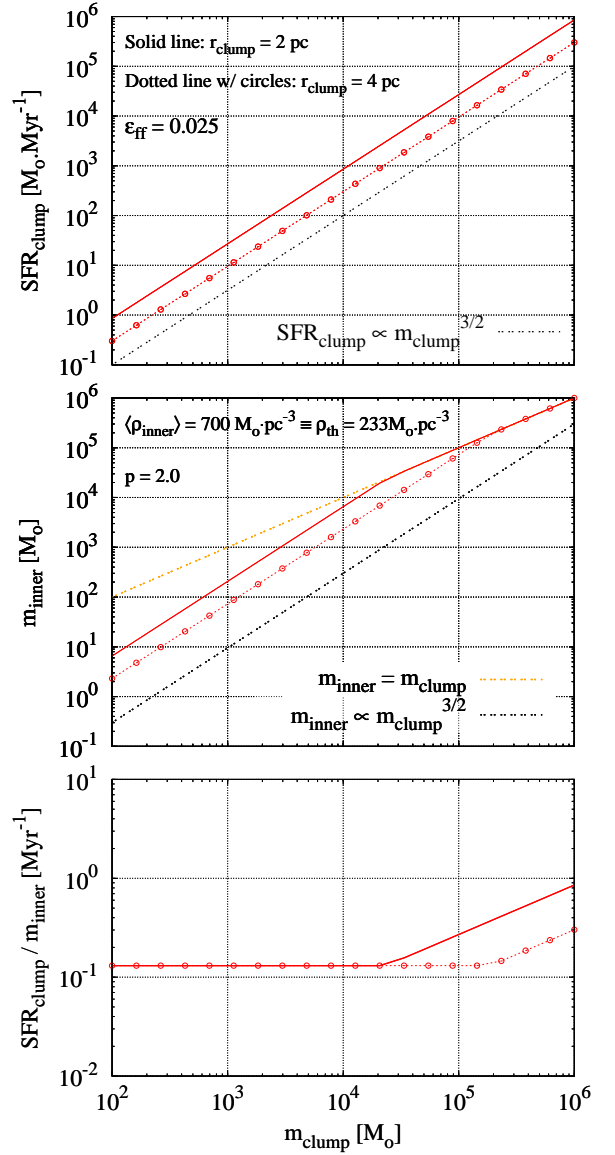}
\caption{{\it Top panel:} Star formation rate of a molecular clump, $SFR_{clump}$, in dependence of its mass, $m_{clump}$.  Parameters are given in the key.  {\it Middle panel:} Mass of the dense inner region of clumps, $m_{inner}$, in dependence of the clump mass.  For a spherically-symmetric clump, the dense inner region is defined such that the gas is denser than a threshold $\rho_{th}=233\,\Msppp$.  The adopted clump density profile is that of an isothermal sphere ($p=2$ in Eq.~\ref{eq:rhoprof}).  The dash-dotted orange line marks $m_{inner} = m_{clump}$.  This regime is met when the clump consists exclusively of gas denser than the adopted threshold.  {\it Bottom panel:} Ratio between the clump \sfr and the mass of its dense inner region in dependence of the clump mass (see text for details)}
\label{fig:insh}
\end{center} 
\end{figure}

Equation \ref{eq:sfrcl} is shown in the top panel of Fig.~\ref{fig:insh} for a \sfe per free-fall time $\eff = 0.025$ and two clump radii, $r_{clump}=2$\,pc (solid red line) and $r_{clump}=4$\,pc (dotted red line with open circles). The black dashed line indicates a logarithmic slope of 3/2. 

Let us assume that the clump density profile is a power-law of the distance $r$ to its center:

\begin{equation}
\rho_{clump}(r) = \rho_{gas,init}(r) \propto r^{-p}\,,
\label{eq:rhoprof}
\end{equation}
with $r \leq r_{clump}$.  We remind the reader that such a density profile does not resolve the dense molecular cores in which individual stars form, and whose number density is of the order of $10^5\,cm^{-3}$ \citep[e.g.][]{hac17}.  Let us divide the clump into two regions based on a density threshold, $\rho_{th}$: {\it (i)} a `dense inner region' where $\rho_{clump}(r) \geq \rho_{th}$, and {\it (ii)} some `low-density outskirts' where $\rho_{clump}(r) < \rho_{th}$.  
The mass of the dense inner region is given by eq.~3 in \citet{par11a}, reproduced here for the sake of clarity:
\begin{equation}
m_{inner} = \left( \frac{3-p}{4 \pi \rho_{th} } \right)^\frac{3-p}{p} m_{clump}^{3/p} \, r_{clump}^{-3(3-p)/p}\;.
\label{eq:mth}
\end{equation}
For this equation to be valid, the density at the clump edge must be lower than the density threshold.  That is, the clump must present some low-density outskirts and $\rho_{edge} = \rho_{clump}(r_{clump}) < \rho_{th}$ or, equivalently, $m_{inner} < m_{clump}$.  
The distance to the clump center corresponding to the density threshold, i.e. $r_{th}$ such that $\rho_{clump}(r_{th})=\rho_{th}$, is given by eq.~4 of \citet{par11a}.   Note that our terminology `inner region' and `outskirts' is applied regardless of their respective spatial extent.

The volume density profile of observed \sfing clumps is often reported to have a slope around $-1.7$ \citep{beu02,mue02,pir09}.  However, the gas density profile is probably steeper at \stf onset since \stf is faster in the clump inner regions whose \fft is shorter \citep[see bottom panel of fig.~1 in][]{par13}.  Let us thus assume that the slope index of the gas density profile is $p=2$ initially.  Equation~\ref{eq:mth} then becomes:
\begin{equation}
m_{inner} = \left( \frac{1}{4 \pi \rho_{th} } \right)^{1/2} \left( \frac{m_{clump}}{r_{clump}} \right)^{3/2}\;.
\label{eq:mdgp2}
\end{equation}
The comparison of Eqs~\ref{eq:sfrcl} and \ref{eq:mdgp2} shows that both the \stf rate of clumps and the mass of their dense inner region scale as $(m_{clump}/r_{clump})^{3/2}$.  As a result, $SFR_{clump}$ and $m_{inner}$ are linearly correlated {\it even though nowhere have we postulated that \stf is limited to the dense inner region of our model clumps}.

To quantify $m_{inner}$, a value of $\rho_{th}$ is still needed.  For a clump density profile with $p=2$, the mean density inside a given radius is three times the density at that radius \citep[see eq.~7 in][]{par11b}.  Given that the mean density of the clumps of \citet{wu05, wu10} is of order $700\Msppp$ (see Section \ref{ssec:dg}), we assume a density threshold $\rho_{th}=233\Msppp$ to define the dense inner region of a molecular clump.  

The middle panel of Fig.~\ref{fig:insh} presents Eq.~\ref{eq:mdgp2} for $\rho_{th}=233\Msppp$, and clump radii identical to those of the top panel.  As for the top panel, the dashed black line indicates a slope of 3/2, that is, the same dependence on the clump mass as that shown by the \stf rate at fixed radius (Eq.~\ref{eq:sfrcl}).  Given that the clump mean density increases along with their mass, the density at the clump edge eventually becomes higher than the adopted density threshold (i.e. $\rho_{edge} > \rho_{th}$).  Equation~\ref{eq:mdgp2} is then no longer valid and must be substituted by $m_{inner} = m_{clump}$ (dashed-orange line in middle panel of Fig.~\ref{fig:insh}), meaning that the clump consists of gas denser than the adopted threshold $\rho_{th}$ only.  There is therefore a break-point beyond which the $(m_{clump},m_{inner})$ relation gets shallower.  

The bottom panel of Fig.~\ref{fig:insh} presents the ratio between the \sfr of the whole clump and the mass of its dense inner region in dependence of the clump mass.  
As long as {\it (i)} the density at the clump edge remains smaller than $\rho_{th}$, {\it (ii)} the \sfe per \fft is constant, and {\it (iii)} $p=2$ in Eq.~\ref{eq:rhoprof}, the \sfr of the {\it whole} clump and the mass of its dense inner region are directly proportional (see Eqs.~\ref{eq:sfrcl} and \ref{eq:mdgp2}):
\begin{equation}
SFR_{clump} = 4 \zeta \eff \sqrt{\frac{2G \rho_{th}}{\pi}} m_{inner}\;.
\label{eq:sfrmdg}
\end{equation}
That is, the ratio $SFR_{clump}/m_{inner}$ is independent of both the clump mass and radius.  At higher clump masses/densities, however, condition {\it (i)} and Eqs~\ref{eq:mth}-\ref{eq:sfrmdg} are no longer valid and $m_{clump}$ scales as $m_{inner}$.  As a result, $SFR_{clump}/m_{inner}$ scales as $m_{clump}^{1/2}$ for a given clump radius \citep[see also][]{kru07b}.  

In the observational parameter space, Eq.~\ref{eq:sfrmdg} translates into a linear correlation between the clump infrared and HCN luminosities.  As we demonstrate in the next section, however, that does not necessarily mean that the dense inner region of clumps is the main site of their star formation activity.

\section{Modeling}
\label{sec:sim}

\subsection{Star mass fraction in clump inner region}
\label{ssec:frac}

In this section, we build on the model of \citet{par13} to compare the total stellar mass of a clump to the stellar mass located in its dense inner region (i.e. where $\rho_{clump}(r) > \rho_{th}$).  

We consider a grid of model clumps whose radius ranges from 0.5\,pc to 8\,pc in logarithmic steps of 0.15, and whose mass ranges from $300\,\Ms$ to $10^6\,\Ms$ in logarithmic steps of 0.25.  All clump density profiles have $p=2$.   Given that the mean number density of HCN(1-0)-mapped gas is usually assumed to be $n_{\rm H2} \simeq 3\cdot 10^4\,cm^{-3}$ \citep{gs04,gb12}, we assign a mean density $\langle \rho_{inner} \rangle = 2100\,\Msppp$ to the clump dense inner region, equivalent  to a density threshold $\rho_{th}=700\Msppp$.  Note that these values are three times higher than those adopted in Section \ref{sec:sfrmth}.  The \sfe per \fft remains $\eff = 0.025$.

\begin{figure}
\begin{center}
\epsscale{1.1}  \plotone{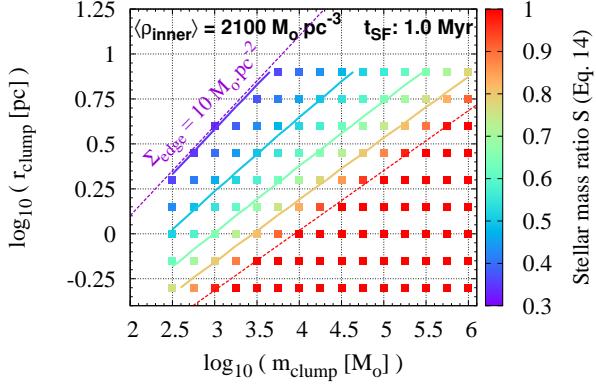}
\caption{Color-coded diagram illustrating the ratio between the stellar mass in the dense inner region of a clump and the total stellar mass formed by that clump, $S=m_{stars,inner}/m_{stars,tot}$.  The ratio $S$ is presented in dependence of  the clump mass ($x$-axis) and radius ($y$-axis), and is color-coded according to the palette at the right-hand side of the plot.  The dashed purple line corresponds to a surface density of $10\,\Mspp$ at the clump edge, and the red dashed line to a mean volume density of $2100\,\Msppp$.  The solid lines in-between highlight the locii of points with a given value of $S$: 0.35 (blue), 0.50 (cyan), 0.65 (light-green) and 0.80 (dark yellow).  Model parameters are: $\eff = 0.025$, $t_{SF}=1$\,Myr and $p$=2.}
\label{fig:map1}
\end{center} 
\end{figure}

Figure \ref{fig:map1} presents the ratio $S$ between the stellar mass formed in the clump inner region, $m_{stars,inner}$, and the clump total stellar mass, $m_{stars,tot}$:
\begin{equation}
S = \frac{m_{stars,inner}}{m_{stars,tot}}
\end{equation}
as a function of the clump total mass ($x$-axis) and radius ($y$-axis).  The time-span since star formation onset is $t_{SF}=1$\,Myr.  Each plain square depicts a model clump and the value of the stellar mass ratio $S$ is indicated by the square color (see the palette at the right-hand side of the plot).  
The upper-left limit of the diagram (purple dashed line) depicts the locus of clumps whose surface density at the edge is $\Sigma_{edge} = 10\,\Mspp$.  This is the gas surface density characterizing the transition between neutral hydrogen and  CO-traced molecular gas \citep{big08}.  The red dashed line corresponds to a mean volume density of $2100\,\Msppp$, hence a density at the clump edge equal to the density threshold, i.e. $\rho_{edge} = \rho_{th}$.  Below that line, clumps are initially made of dense gas only, and the stellar mass ratio $S$ is thus systematically equal to unity.  The intermediate solid lines highlight the locii of points with a given $S$ ratio, from top to bottom: $S=0.35$ (blue), $S=0.50$ (cyan), $S=0.65$ (light-green) and $S=0.80$ (dark yellow).  For instance, a clump with a mass of $10^4\,\Ms$ and a radius $\simeq 4$\,pc has 50\,\% of its stellar mass located at $\rho_{clump(r)} > \rho_{th}$.  Clumps with a mean density lower than $40\,\Msppp$, which roughly corresponds to the region above the cyan line ($S=0.50$), have a stellar mass fraction at $\rho_{clump(r)} > \rho_{th}$ systematically lower than 50\,\%.  That is, their low-density outskirts have been more active at forming stars than their dense inner region.  This occurs despite the volume-density-driven nature of the model implemented here.  

\begin{figure}
\begin{center}
\epsscale{1.1}  \plotone{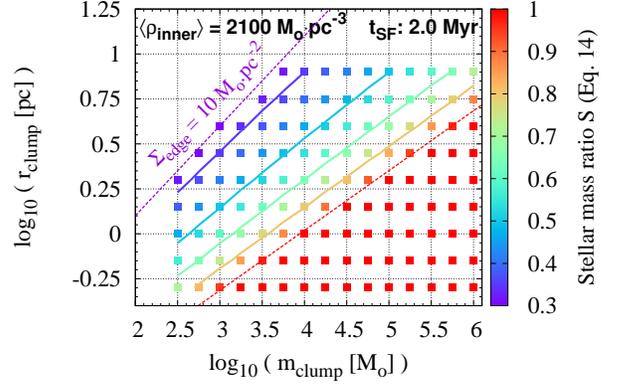}
\caption{Same as Fig.~\ref{fig:map1} but for $t_{SF} = 2$\,Myr}
\label{fig:map2}
\end{center} 
\end{figure}

As time goes by, that trend gets reinforced.  Figure~\ref{fig:map2} displays the same information as Fig.~\ref{fig:map1} at a time $t_{SF} = 2$\,Myr after \stf onset.  As one can see,
clumps of a given total mass and radius are now characterized by a lower fraction of stars in their dense inner region than at $t_{SF}=1$\,Myr (i.e. the solid lines have moved downwards).  This effect stems from the \sfr dropping more slowly in the clump outer regions than in the inner regions because of their different free-fall times.  This effect is also noticeable in the evolution with time of the embedded-cluster density profile, which gets slightly shallower with time \citep[see fig.~5 in][]{par13}.  

Now following the analytical insights of Section \ref{sec:sfrmth}, let us investigate how well the total \sfr of the model clumps correlate with their dense-gas mass at a given time $t_{SF}$ after star formation onset.   

\subsection{Dense-gas mass versus total SFR correlation}
\label{ssec:cor}

\begin{figure}
\begin{center}
\epsscale{1.1}  \plotone{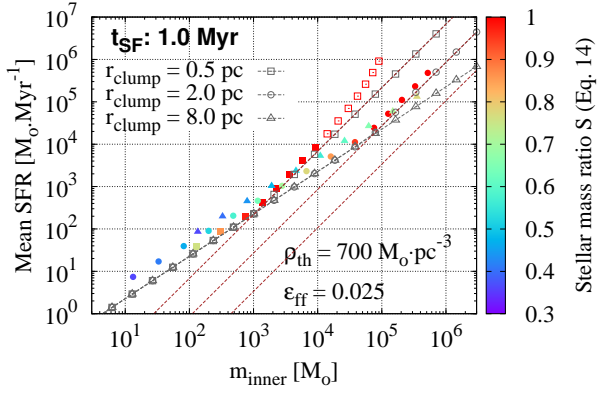}
\caption{Correlation between the mass of the dense inner region and the mean \stf rate of cluster-forming clumps.  Colored symbols correspond to model predictions, with the same color coding and model parameters as in Fig.~\ref{fig:map1}.  Squares, circles and triangles correspond to clump radii of 0.5, 2.0 and 8.0\,pc.  Plain/open symbols correspond to global \stf efficiencies $SFE_{gl}$ lower/higher than 50\,\%.  Grey lines with open symbols correspond to the analytical predictions of Section~\ref{ssec:no}.  Dashed brown lines depict, for each clump radius, the scaling expected when the clump is made of dense gas only, that is, $SFR \propto m_{dg}^{3/2}$ (see Eq.~\ref{eq:sfrcl})} 
\label{fig:cor}
\end{center} 
\end{figure}

For the same model clumps as in Fig.~\ref{fig:map1}, Fig.~\ref{fig:cor} shows the relation between the mass of the dense inner region and the mean \stf rate at $t_{SF} = 1$\,Myr.  We define the mean \sfr as the ratio between the clump total stellar mass and the corresponding \stf timespan:
\begin{equation}
SFR_{mean} = \frac{m_{stars,tot}}{t_{SF}}\;.
\end{equation}
The color-coding is identical to that used in Fig.~\ref{fig:map1}.  The squares, circles and triangles correspond to clumps with radii of 0.5, 2.0 and 8.0\,pc, respectively.  [Additional explanations about symbol coding follow later in this section].  The grey lines with open squares, circles and triangles depict the analytical models presented in Section~\ref{ssec:no}, with $\rho_{th} = 700\,\Msppp$.  

The offset between the analytical predictions (grey lines) and the models (individual colored symbols) is driven by several factors.  One is the decrease of the inner-region gas mass, which Eq.~\ref{eq:mth} does not account for.  The second is the decrease with time of the \stf rate.  This is a characteristic of the cluster-formation model implemented here, which stems from the gas \fft becoming longer as gas feeds star formation.  It is especially noticeable in high-density, hence short free-fall time, environments \citep[see fig.~3 in ][and Fig.~\ref{fig:msfrwtime} in Section \ref{ssec:corwtime}]{par14}.  

At this stage, we note that the mass of the clump inner region underestimates the mass in clump dense gas.  
In the L1495/B213 Taurus region and the Perseus NGC1333 protocluster, most Class~0/I objects are observed in `fibers', also refered as `velocity-coherent filaments' \citep{hac13,hac17}.  Their formation may result from instabilities affecting the molecular gas at densities lower than $10^4\,{\rm cm}^{-3}$ \citep[][their section 7.5]{hac13}.  Such fibers have been detected in $N_2H^+$ \citep[in the L1495/B213 Taurus region, ][]{hac13} and in $N_2H^+$ and $NH_3$ \citep[in the Perseus NGC1333 protocluster, ][]{hac17}.  Given that $N_2H^+$ and $NH_3$ trace gas with a number density higher than $5 \cdot 10^4 {\rm cm}^{-3}$ \citep{hac17}, such fibers contribute to the HCN luminosity and dense-gas mass of a \sfing region whose mean number density is nevertheless lower than $10^4\,cm^{-3}$.  Therefore, the total dense-gas mass of a model clump, $m_{dg}$, amounts to the mass of its dense inner region plus the mass in fibers present in its low-density outskirts:
\begin{equation}
m_{dg} = m_{inner} + m_{fibers, outskirts}.
\label{eq:fib1}
\end{equation}
\citet{hac17} find that, in NGC1333, the mass in fibers is ten times the mass in Class~0/I objects, regardless of the location inside the protocluster.  To estimate $m_{fibers, outskirts}$, we thus derive the stellar mass formed in the outskirts of our model clumps over the past 0.5\,Myr \citep[i.e. the lifespan of Class~I objects, ][]{eva09}, $m_{Cl0/I, outskirts}$.  We then obtain the total dense-gas mass of a clump, $m_{dg}$, as:
\begin{equation}
m_{dg} = m_{inner} + f \cdot m_{Cl0/I, outskirts}\,,
\label{eq:fib2}
\end{equation}   
with $f=10$.  We note that because the detection density threshold of $N_2H^+$ is slightly higher than that of HCN (i.e. $ n_{th, N_2H^+} \simeq 5 \cdot 10^4 {\rm cm}^{-3}$ vs. $ n_{th, HCN} \simeq 10^4 {\rm cm}^{-3}$) and, since we define the dense gas as HCN-mapped gas, the actual value of $f$ in Eq.~\ref{eq:fib2} may be slightly higher than $10$.  

\begin{figure}
\begin{center}
\epsscale{1.1}  \plotone{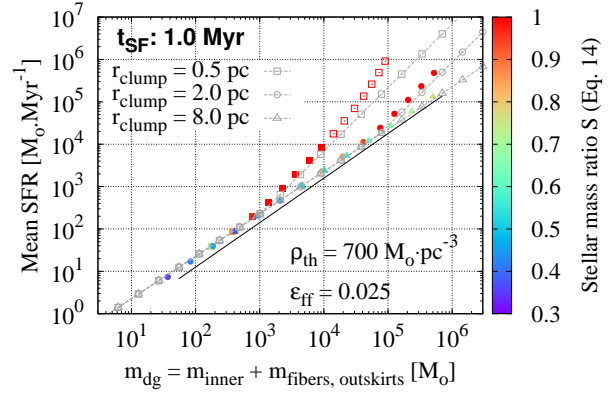}
\caption{Correlation between the total mass of dense gas and the mean \stf rate of the same model clumps as in Fig.~\ref{fig:cor}.  The dense-gas mass of a clump, $m_{dg}$, is defined as the mass of its dense inner region, $m_{inner}$, increased by the mass in dense fragments seeding the formation of Class~0/I objects in its low-density outskirts, $m_{fibers, outskirts}$.  Coloured symbols correspond to $f=10$ in Eq.~\ref{eq:fib2}.  The solid black line depicts where the model clumps with $S<1.0$ (i.e. clumps with low-density outskirts, depicted in this figure as non-red points) move when $f=20$.  Grey lines with open symbols as in Fig.~\ref{fig:cor}  
} 
\label{fig:wfib}
\end{center} 
\end{figure}

Figure~\ref{fig:wfib} presents the total \sfr of our model clumps in dependence of their total dense-gas mass, as defined by Eq.~\ref{eq:fib2} with $f=10$.  Adding the contribution of the fibers strengthens the linearity of the \stf relation in the low-density regime, that is, for the clumps whose mean density $<2100\,\Msppp$ (compare the non-red symbols in Fig.~\ref{fig:cor} and Fig.~\ref{fig:wfib}).  For the sake of completeness, Fig.~\ref{fig:wfib} also shows as a solid black line their \stf relation when $f=20$.      


The star formation relations of Fig.~\ref{fig:wfib} behave linearly as long as $m_{dg} < 10^4\,\Ms$.  For larger dense-gas masses, they steepen and scale as $SFR \propto m_{dg}^{3/2}$, rather than $SFR \propto m_{dg}$.  For a given clump radius, the break-point corresponds to the density at the clump edge equating the adopted density threshold, i.e. $\rho_{edge} =  \rho_{th}$.  Beyond the break-point, the whole clump consists of gas with a mean density higher than $2100\,\Msppp$ and a \fft shorter than 0.18\,Myr.  This yields a rise of the \stf rate with the dense gas mass steeper than below the break-point.  
\citet{kru07b} predict the same effect for the $\lhcn - \lir$ relation based on their modeling of galaxy regions with volumes of $10^7$, $10^8$, and $10^9$\,pc$^3$ (corresponding radii of $\simeq$ 135, 300, and 600\,pc).  Specifically, their model of the $\lhcn - \lir$ (or $L_{\rm HCO+} - \lir$) relation steepens from a slope of unity to a slope of 3/2 once the mean gas density becomes higher than the critical density of the molecular line tracer.  The origin of the slope steepening is thus similar in both theirs and the present model, that is, a change in the mean density of the dense/mapped gas.

We note that our results for model clumps beyond the break-point must be considered with caution because $t_{SF}$ may be exceedingly large compared to the clump initial free-fall time.  For instance, the initial \fft of our densest model clump is $\simeq 6000$\,yrs (initial gas mass of $10^6\,\Ms$ and radius of $0.5$\,pc).  A time-span of $t_{SF} = 1$\,Myr thus corresponds to 170 free-fall times.  Clearly, a static model like the one of \citet{par13} cannot handle the evolution of a clump for that many free-fall times.  Besides, the clump global \stf efficiency, $SFE_{gl}$, namely the ratio between the total stellar mass formed by the clump and its initial gas mass

\begin{equation}
SFE_{gl} = \frac{m_{stars,tot}}{m_{clump}}
\end{equation}

 becomes exceedingly large.  For instance, the global \sfe predicted for the densest model clump is $SFE_{gl} \simeq 0.9$, as Fig.~\ref{fig:wfib} shows that the residual gas mass at $t_{SF}=1$\,Myr is $\simeq 10^5\,\Ms$ (see the top red open square).  To achieve such an extreme global \sfe is highly unlikely since the massive stars would certainly expel the residual gas at an earlier time.  Let us assume that embedded clusters blow out their residual gas and become visible in the optical once they have achieved a global \sfe $SFE_{gl}=0.50$.  How does the removal of those now exposed star clusters from Fig.~\ref{fig:wfib} affect the shape of the \stf relation?  In Fig.~\ref{fig:wfib}, clusters with $SFE_{gl} < 50$\,\% are marked with filled symbols, while those with $SFE_{gl} \geq 50$ are depicted by open symbols.  Removing clusters with $SFE_{gl} > 0.50$ from Fig.~\ref{fig:wfib} (i.e. open-symbols are now discarded) strongly reduces the upturn characterizing the \stf relation when $m_{dg} > 10^4\,\Ms$.  We have to keep in mind, however, that that region may still be occupied by younger clumps, i.e. clumps with a \sfe lower than 50\,\% at $t_{SF} < 1$\,Myr.  We return to this point in Section \ref{ssec:corwtime}.  

\begin{figure}
\begin{center}
\epsscale{1.1}  \plotone{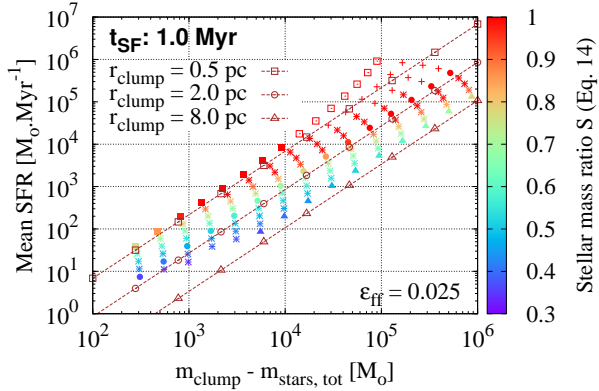}
\caption{Correlation between the residual gas mass and mean \stf rate of the model clumps shown in Fig.~\ref{fig:map1}.  The brown lines with open symbols depict, for each clump radius, the analytical models presented in Section \ref{ssec:no} (see Eq.~\ref{eq:sfrcl}).  They are identical to the dashed symbol-free brown lines of Figs~\ref{fig:cor}.  Squares, circles and triangles correspond to clump radii of 0.5, 2.0 and 8.0\,pc, respectively.  Asterisks and plusses correspond to all other clump radii.  Plain symbols and asterisks correspond to global \stf efficiencies lower than 50\,\%.  Open symbols and plusses correspond to global \stf efficiencies higher than 50\,\%.  
}
\label{fig:cor2}
\end{center} 
\end{figure}

Figure \ref{fig:cor2} shows the mean \sfr in dependence of the clump residual gas mass, i.e. $m_{clump} - m_{stars,tot}$, at $t_{SF}=1$\,Myr.  Although both quantities are correlated, the distribution of points is markedly wider than in Fig.~\ref{fig:wfib}.  Figures~\ref{fig:wfib} and \ref{fig:cor2} can also be qualitatively compared to fig.~1 in \citet{lad12}, which shows the mean \sfr of nearby molecular clouds as a function of their total gas- and dense-gas masses.  As emphasized by \citet{lad12}, nearby clouds also show a correlation between dense-gas mass and \sfr tighter than between total gas mass and \stf rate.  We refrain from doing a more quantitative comparison, however, since \citet{lad12} define the dense gas content of nearby clouds based on an extinction threshold ($A_{K, th} = 0.8$\,mag, equivalent to a gas surface density of $\simeq 160\,\Mspp$).  There is no certainty that the dense-gas mass estimates 
defined with a gas surface density criterion would always match those defined with a volume density threshold.  
In a forthcoming paper, we will derive the \stf relations predicted by our model when using a surface-density threshold criterion, and compare them to those obtained in this contribution. \\

\subsection{Star formation relation and star formation efficiency of the fibers}
\label{ssec:sfrap}

Some of our model clumps have diameters as large as 16\,pc.  When plotting the \sfr of a clump as a function of its dense-gas mass (see Fig.~\ref{fig:wfib}), we nevertheless  implicitly assume that each clump is mapped in its entirety.  What \stf relation would we infer if the clumps were to be oberved with an aperture smaller than their diameter, and covering only a fraction of the clump outskirts.  The dense fibers observed in the aperture are then the only contributors to the dense-gas mass, $m_{dg,aperture}$, which can be estimated from the corresponding number of Class~0/I objects, $N_{Cl0/I, aperture}$ (see Section~\ref{ssec:cor}): 
\begin{equation}
m_{dg,aperture} = m_{fibers,aperture} = f \cdot N_{Cl0/I, aperture} \cdot 0.5\,\Ms\;.
\end{equation}   
In this equation, 0.5\,$\Ms$ is the mean stellar mass of Class~0/I objects, and $f$ is the ratio between the masses in fibers and in Class~0/I objects (see Eq.~\ref{eq:fib2}).  
As for the measured \stf rate, it can be derived from the number of Class~0/I objects combined to their lifespan \citep[0.5\,Myr, ][]{eva09}:
\begin{equation}
SFR_{aperture} = \frac{N_{Cl0/I} \cdot 0.5\,\Ms}{0.5\,Myr}.
\end{equation}

The ratio of both equations yields the relation between the \sfr and the dense-gas mass expected for an aperture covering a fraction of the clump outskirts:
\begin{equation}
SFR_{aperture} = 0.2 m_{dg,aperture}
\label{eq:sfr_ap}
\end{equation}
where we have adopted $f=10$.  Eq.~\ref{eq:sfr_ap} is remarkably close to our model \stf relation (see all non-red symbols in Fig.~\ref{fig:wfib}).  We therefore deduce that an aperture covering part of the clump outskirts, whose fibers constitute the only contribution to the dense gas mass, still yields the same \stf relation as that shown by our models in Fig.~\ref{fig:wfib}.  Since we consider the dense gas only, reducing the aperture to the point that it covers a dense-gas fragment and its corresponding Class~0/I objects does not modify the inferred \stf relation.

\subsection{Star formation relations and clump evolution}
\label{ssec:corwtime}

\begin{figure}
\begin{center}
\epsscale{1.1}  \plotone{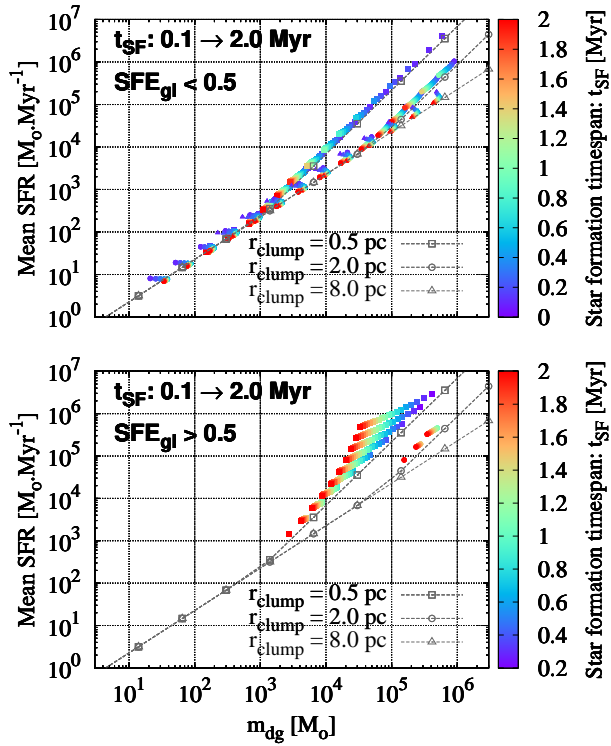}
\caption{Mean \sfr in dependence of the dense-gas mass for molecular clumps with radii of 0.5, 2.0 and 8.0\,pc, from $t_{SF}=0.1$\,Myr to $t_{SF}=2$\,Myr in steps of 0.1\,Myr.  The top and bottom panels show clumps with global \stf efficiencies lower and higher than $SFE_{gl} = 0.5$, respectively.  The color-coding refers to the timespan since star-formation onset, $t_{SF}$.  The grey lines with open squares, circles and triangles depict the analytical models presented in Section~\ref{ssec:no}, with $\rho_{th} = 700\,\Msppp$, and already shown in Figs~\ref{fig:cor} and \ref{fig:wfib}.}
\label{fig:msfrwtime}
\end{center} 
\end{figure}

Figure \ref{fig:msfrwtime} shows the mean \sfr in dependence of the dense-gas mass for molecular clumps with radii of 0.5, 2 and 8\,pc, from $t_{SF}=0.1$\,Myr to $t_{SF}=2$\,Myr in steps of 0.1\,Myr.  It thus provides a more complete picture than Fig.~\ref{fig:wfib}, which presents a snapshot at $t_{SF} = 1$\,Myr only.  Note that the color-coding of Fig.~\ref{fig:msfrwtime} refers to the timespan since the onset of star formation, $t_{SF}$, and not to the stellar mass ratio $S$ as done for Figs~\ref{fig:cor} and \ref{fig:wfib}.  For the sake of clarity, the results are displayed in two distinct panels depending on the clump global \sfe (top panel: $SFE_{gl} \leq 0.50$; bottom panel: $SFE_{gl}>0.50$).  
 
As already noticed in Figs~\ref{fig:cor} and \ref{fig:wfib}, the upturn (i.e. when the $m_{dg}-SFR$ relation becomes steeper than linear) is populated by clumps with $SFE_{gl}>0.50$ (see bottom panel).  The top panel, however, shows it to be also populated by clumps with $SFE_{gl}\leq0.50$.  

If there is a critical \sfe beyond which clumps are no longer observed as a result of stellar feedback having expelled the residual \stf gas (say, $SFE_{gl} = 0.5$ as assumed in Sec.~\ref{ssec:cor}), there must be a corresponding limit for the \stf timespan.  Inspection of the top panel of Fig.~\ref{fig:msfrwtime} actually reveals that the majority of the clumps with $SFE_{gl} < 0.5$ and populating the upturn have $t_{SF} \leq 1$\,Myr (the exact value depends on the choice of $\eff$).  This fairly limited timespan stems directly from the high mean density of the clumps populating the upturn ($\geq 2100\,\Ms \cdot pc^{-3}$) since, as shown in \citet{kru07b} and Sec.~\ref{ssec:no}, the upturn is populated by \sfing regions whose density at the edge is higher than the adopted density threshold (i.e. $\rho_{edge} > \rho_{th}$).  As a result, clumps populating the upturn evolve faster than those populating the linear segment of the  $m_{dg}-SFR$ relation.  Therefore, if we assume that clumps with a global \sfe higher than 0.5 are no longer observed in HCN emission, 
our results suggest that the dense cluster-forming clumps populating the upturn experience shorter \stf episods than those lying on the linear part of the $m_{dg}-SFR$ relation.

In the framework of our model, we may also expect the $\lhcn - \lir$ relation to be superlinear rather than linear (or to show a double-index power-law).  This cannot be tested with the samples of \citet{wu05} and \citet{lad12}, due to their poor coverage of the $m_{dg} \gtrsim 10^4\,\Ms$ regime \citep[only a few such clumps are included in the sample of][]{wu05}.  
Yet, \citet{gb12} find a superlinear relation between the HCN and infrared luminosities of galaxy regions (size of a few kpc) observed in a sample of normal \sfing galaxies, LIRGS and ULIRGS \citep[their fig.~3; see also][]{gao07}.  Their galaxy sample is larger than that for which \citet{gs04} initially found a linear $\lhcn -\lir$ relation \citep[see also][]{gao07}.  Note that prior to comparing our model predictions to such observations, one still needs the evolution with time of the infrared luminosity of cluster-forming clumps whose star formation rate decreases with time, as given by eq.~14 of \citet{par14}.

\section{Conclusions}
\label{sec:conclu}

In this contribution, we have modeled the \stf relation linking the dense-gas mass and \sfr of spherical molecular clumps whose volume density profile scales as $\rho_{clump}(r) \propto r^{-2}$ ($r$ is the distance to the clump center).  We consider as dense gas gas whose density is higher than a threshold $\rho_{th} = 700\,\Ms \equiv n_{\rm H2, th} = 10^4\,cm^{-3}$.  Based on this, we divide each model clump into 2 regions: a dense inner region where $\rho_{clump}(r) > \rho_{th}$, and the low-density outskirts where $\rho_{clump}(r) < \rho_{th}$.  To derive the clump \stf rate, we assume a constant \sfe per \fft  \citep{kru05}.         
Our results are as follow:

1.~The {\it total} \sfr of the clumps scales linearly with the mass of their dense inner region.  This result holds even though the \sfr considered here is that of the whole clump, and not that of the dense inner region only.  It has been established both via analytical insights (Section~\ref{ssec:no}, Eq.~\ref{eq:sfrmdg}), and also based on the model of \citet{par13} (Section~\ref{sec:sim}). The linear scaling between total \sfr and clump inner-region mass stems from them depending in the same way on the clump mass and radius (see Eqs~\ref{eq:sfrcl} and \ref{eq:mdgp2}).  This is true (i) as long as the gas initial density profile scales as $\rho_{clump}(r) \propto r^{-2}$, and (ii) the clump has some low-density outskirts with $\rho_{clump}(r) < \rho_{th}$.  When the volume density at the clump edge is higher than the density threshold (i.e. $\rho_{edge} > \rho_{th}$: the clump is made of dense gas only), the relation between dense gas mass and \sfr becomes superlinear (slope = 3/2).  This is because as the clump becomes denser, its \fft gets shorter, thereby promoting a steeper rise of the \stf rate than would be predicted by simply extending what has been found at lower densities \citep[see also][]{kru07}.  

2.~We have estimated what fraction of their stellar mass clumps actually form in their dense inner region.  To do so, we have applied the model of \citet{par13} to a grid of model clumps, with a radius ranging from 0.5\,pc to 8\,pc and a mass ranging from $300\,\Ms$ to $10^6\,\Ms$.  The results are presented in Figs~\ref{fig:map1}-\ref{fig:map2}. The stellar mass fraction in the dense inner region varies from unity, when clumps are made  of dense gas only, down to $\simeq 1/3$, when the surface density at the clump edge is $\Sigma_{edge} = 10\,\Mspp$ (i.e. the transition between neutral hydrogen and CO-traced molecular gas).
Therefore, clumps can yield a linear \stf relation in terms of the dense inner-region mass even if more than half of their stellar mass form {\it out} of that region.  This is seen by comparing Figs \ref{fig:map1} and \ref{fig:wfib}.   {\it We therefore emphasize that the linearity of a \stf relation should not be straightforwardly interpreted as evidence that \stf takes place exclusively in the gas whose mass is given as the input parameter of the \stf relation}.  In the case under scrutiny here, the linear \stf relation between the mass of the dense inner region and the \sfr does not imply that the \stf activity is that of the dense inner region.  This is the total \stf activity of the clump that we have considered.   
  
3.~To improve the accuracy of our results, we have accounted for the mass in dense-gas fragments seeding \stf in the clump low-density outskirts (see Section~\ref{ssec:cor}).  These fragments are not resolved by the adopted clump density profile, $\rho_{clump}(r)$.  To do so, we use the ratio 
between the mass of $N_2H^+$-mapped gas and the mass in Class~0/I objects in the Taurus L1495/B213 region, which is found to be around $10$ by \citet{hac17}.  Applying this correction (Eq.~\ref{eq:fib2}) strengthens the linearity of the dense-gas mass vs. \sfr relation for clumps with $\rho_{edge} < \rho_{th}$.    %
  
4.~We have obtained how the dense-gas mass and \sfr of our model clumps evolve with time (Fig.~\ref{fig:msfrwtime}).  We suggest that the clumps populating the upturn of the \stf relation are younger than those lying on its linear segment.  This is because the clumps on the upturn have no low-density outskirts (i.e. $\rho_{edge} > \rho_{th}$), have thus a higher mean density and reach on a faster time-scale (in units of Myr) the critical \sfe beyond which stellar feedback expels the clump residual gas \citep[see also][]{par14}.  We suggest the dual behavior of our model \stf relation as a possible explanation for the superlinear $\lhcn - \lir$ relation found by \citet{gb12} for galactic regions with sizes a few kpc.  

5.~We have compared the \stf relation in dependence of the clump dense gas with the \stf relation in dependence of the clump total gas (Figs~\ref{fig:wfib} and \ref{fig:cor2}, respectively).  The former is significantly tighter than the later.  This effect has been observed for molecular clouds of the \Son by \citet{lad12} and \citet{eva14}, and for more distant clouds of the Galactic plane by \citet{vut16}. 

6.~While individual stars form in fragments/cores made of dense gas, as suggested by the linear $\lhcn - \lir$ relation, we caution that that same relation does not imply that \sfing regions consist of dense gas only when the density is averaged on a pc-scale.



\acknowledgments
GP thanks the referee for a very constructive report, and Alvaro Hacar for interesting discussions during the `VIALACTEA2016' conference (Rome, September 2016).  
She acknowledges support from the Sonderforschungsbereich SFB 881 "The Milky Way System" (subproject B2) of the German Research Foundation (DFG).







\end{document}